\documentclass[preprint,showpacs,preprintnumbers,amsmath,amssymb]{revtex4}
\usepackage{graphicx} 
\usepackage{dcolumn} 
\usepackage{bm} 

\begin{document}

\title{Bipolar single-wall carbon nanotube field-effect transistor}

\author{Bakir.\ Babi\'c}
\author{Mahdi.\ Iqbal}
\author{Christian.\ Sch{\"o}nenberger}
\email{Christian.Schoenenberger@unibas.ch}
\homepage{www.unibas.ch/phys-meso} \affiliation{Institut f\"ur
Physik, Universit\"at Basel, Klingelbergstr.~82, CH-4056 Basel,
Switzerland }
\date{\today}

\begin{abstract}
We use a simultaneous flow of ethylene and hydrogen gases to grow
single wall carbon nanotubes by chemical vapor deposition. Strong
coupling to the gate is inferred from transport measurements for
both metallic and semiconducting tubes. At low-temperatures, our
samples act as single-electron transistors where the transport
mechanism is mainly governed by Coulomb blockade. The measurements
reveal very rich quantized energy level spectra spanning from
valence to conduction band. The Coulomb diamonds have similar
addition energies on both sides of the semiconducting gap.
Signatures of subbands population has been observed.

\end{abstract}


\maketitle


Single wall carbon nanotubes (SWNTs) are chemically derived
self-assembled solid molecules with fascinating electronic
properties \cite{Dresselhaus}. Their rich variety of band
structure (metallic, semiconductor) might revolutionize
nano-electronics. Recently, not only an extensive spectrum of
quantum phenomena have been demonstrated with SWNTs \cite{Dekker1,
Dai2,mcuene1,Liang}, but they have also been used as functional
electronic devices in the form of field effect transistors (FETs)
\cite{Tans1,Martel1}. The as-grown SWNT FETs were found to be
unipolar p-type, i.e. no electrical current flows even at large
positive gate voltages. The p-type nature of nanotubes (NTs) has
been attributed to charge transfer caused by either oxidizing
molecules adsorbed to the NTs \cite{Dai1}, or the difference in
workfunctions between NTs and metallic contacts (mostly Au)
\cite{Tans1}. The unipolarity, on the other hand, has been
attributed to the presence of Schottky barriers at the
metal-nanotube contacts \cite{Park}. Though as-grown SWNTs are
p-type, n-type unipolar conductance has been demonstrated by
either chemical doping \cite{Bockrath} or an annealing treatment
in an inert environment \cite{Radosavljevic}. It remains, however,
challenging to realize bipolar SWNT FETs which operate without any
additional treatment and use conventional back-gating. Bipolar
SWNT FETs has been demonstrated on large-diameter SWNTs
\cite{Javey}, and recently also on small-diameter SWNTs by using
strong-coupling gates \cite{Bachtold}. Here we report on
electronic transport measurements on as grown SWNTs which show
bipolar FETs action.

SWNTs are synthesized by chemical vapor deposition (CVD) following
the method of Hafner et al. \cite{Hafner1}. In all our studies we
used SWNTs having diameters of \mbox{$2$\,nm} or less, as inferred
from AFM height measurements. Our devices are prepared on highly
doped \mbox{($<$\,0.02\,$\Omega$cm)} and thermally oxidized
\mbox{($400$\,nm)} Si wafers. The substrate is used as back gate
in electrical measurements of the final devices which are obtained
as follows: The substrate is covered with a layer of
polymethylmethacrylate (PMMA) in which windows are patterned by
electron beam lithography. Then, a catalyst suspension consisting
of \mbox{$1$\,mg} iron nitrate seeds (Fe(NO$_3$)$_3$) dissolved in
\mbox{10\,ml} of isopropanol is poured into the predefined
trenches. The PMMA is then removed in acetone, leaving isolated
catalyst islands \mbox{($5\times 10$\,$\mu m^2$)} on the surface.
The CVD growth is performed in a quartz-tube furnace at
\mbox{$800$\,$^\circ$C} and atmospheric pressure using a gas
mixture of ethylene, hydrogen and argon with respective flow rates
of $2$, $400$, and \mbox{$600$\,cm$^3$/min}. During heating and
cooling of the furnace, the quartz tube is continually flashed
with argon to avoid contamination of the tubes. The as-grown SWNTs
are then contacted in a conventional lift-off process with two
metal electrodes per SWNT, spaced  \mbox{$1$\,$\mu$m} apart. As
electrode material Ti/Au bilayer is used, leading to contact
resistances of \mbox{$\approx 40$\,k$\Omega$} at room temperature.
Figure~1a illustrates schematically a SWNT device. An atomic force
microscopy (AFM) picture is displayed in figure~1b. The diameter
of the nanotubes is determined from the measured height using AFM
in tapping mode.

Once the samples are made, semiconducting and metallic tubes are
distinguished by the dependence of their electrical conductance
$G$ on the gate voltage $V_g$, measured in a wide temperature
range of \mbox{$0.3 \dots 300$\,K}. Figure~2 shows $G(V_g)$ for a
semiconducting SWNT at moderate temperatures of \mbox{$T = 40$}
and \mbox{$60$\,K}. Starting from \mbox{$V_g = -10$\,V}, $G$
decreases with increasing $V_g$ indicating p-type behavior, while
above \mbox{$V_g \approx 4$\,V} $G$ increases indicating n-type
behavior. In between these two regions the conductance is low,
which suggests carrier depletion. This low conductance region
corresponds therefore to the gap. The charge-neutrality point for
this sample lies at \mbox{$V_g=2.5$\,V}. Taking the respective
capacitances into account (see below), this corresponds to a Fermi
energy of \mbox{$E_F=0.5$\,eV}. In general, charge-neutrality
points vary between \mbox{$E_F=-0.5$\,eV} and \mbox{$0.5$\,eV}.

Our finding demonstrates that the SWNT devices are bipolar
transistors. Taking a linear approximation for $G(V_g)$ (dashed
lines in figure~2), we obtain a relatively high average carrier
mobility of \mbox{$\mu \approx 800$\, cm$^2$/Vs}.

$G(V_g)$ is not strictly linear, but shows several pronounced
humps (see arrows in figure~2) which we attribute to van Hove
singularities (VHS) in the $1$-dimensional (1D) density-of-states
(DOS). Conductance peaks are expected, if the contacts couple
weakly to the NTs (tunnelling contacts) and if the band structure
of the NT can rigidly move while sweeping $V_g$. The measured low
conductance \mbox{$\approx 10^{-3}\ \times\ 2e^2/h$} of this
device is in favor of tunnelling contacts.  The observed humps in
$G(V_g)$ are separated by \mbox{$\Delta V_g\approx 2$\,V}. This
relates to an energy interval of \mbox{$\approx 0.4$\,eV}. The VHS
are smeared and do not appear symmetrically with respect to the
semiconducting gap. We attribute this to defects which modify the
band structure, resulting in a broadening of the VHS.

We now turn to low-temperature measurements $T \leq 4$ K. Figure~3
shows $G(V_g)$ and a greyscale representation (inset) of the
differential conductance $dI/dV$ as a function of $V_g$ and
applied transport voltage $V_{sd}$ of a SWNT device at \mbox{$T =
2$\,K}. The large white zone in the middle of the greyscale plot
corresponds to a non conducting region related to the
semiconducting gap (SG). The drawn thick lines at the edges are
guides to the eye. Their vertical extensions intersect around
\mbox{$V_{sd} \cong 0.6$\, eV}, which is a direct estimate of the
gap energy. On both sides of the SG  Coulomb blockade  diamonds
(CBD) of varying size are observed (we refer to the term
`diamond', although the blockade region is not composed of a
series of neat diamonds). Though the addition energy $E_{add}$ is
seen to fluctuate in between \mbox{$2.5$} and \mbox{$\leq
20$\,meV}, there is a general trend, indicated by the thin curved
lines. Close to the gap $E_{add}$ is large and decays to smaller
value for lower (higher) $V_g$ on the p (n) side. As a reference
we expect \mbox{$E_{add}\sim 4$\,meV} for an undisturbed
\mbox{$1$\,$\mu$m} long SWNT. Because $E_{add}$ is determined by
the sum of the single-electron charging energy $U_c$ and the $0$D
level spacing $\Delta E$, $U_c$ and $\Delta E$ must depend on
$V_g$, at least in the vicinity of the SG. This is not expected
for an ideal (defect-free) semiconducting NT for the following
reason: At the onset of the conduction or valence band, the $1$D
DOS is expected to be very large (VHS), since the band dispersion
is parabolic to first approximation. If the nanotube can be
considered a single quantum dot extending from one contact to the
other, the $0$D level spacing should be very small, i.e.
\mbox{$\Delta E=0$} to first approximation. Provided the added
charge can spread homogeneously along the whole tube, a constant
charging energy $U_c$ is expected. Hence, we would expect a
constant addition energy in case of an ideal defect-free tube. The
observed discrepancy can be resolved if (weak) disorder is taken
into account. Disorder will distribute the states over some energy
interval leading to the observed broadening of the VHS. Moreover,
this results in a smooth onset of the DOS and consequently in a
relatively large $0$D level-spacing. Disorder also (partially)
localizes the wavefunctions, leading to both increased $\Delta E$
and $U_c$.

Next we focus on the region far away from the SG where ideally the
$0$D wavefunctions are extended, i.e. one quantum dot (QD), and
where the constant interaction model should yield a good
approximation to single-electron charging effects. The charging
energy is then given by the capacitances determined by the
geometry of the device. Figure 4 presents three greyscale plots of
the differential conductance $dI/dV$, 4a and b for one
semiconducting NT and 4c for another metallic NT with identical
contact spacing. The latter is shown as reference. Figure~4a and b
correspond to conduction in the p and n region, respectively. Both
have been measured around \mbox{$|V_g|\approx 10$\,V}
corresponding to $\approx 1000$ added carriers.\cite{us1} If we
compare with our metallic reference, the Coulomb blockade (CB)
diamonds are less regular in the semiconducting case. Importantly,
however, they are similar in all respects on the n and p side. The
irregularity of the CB and the fact that the gap most often does
not close implies that the NT does not act as a single QD. The low
maximum conductance of \mbox{$\leq 0.025$\,$e^2/h$} observed in
the linear conductance in figure~3 supports this finding. Based on
the maximum $E_{add}$, three QD's is an upper limit. Hence, even
at large gate voltage, defects are still effective in dividing the
tube into smaller segments. The CB-pattern on the n-side seems to
show a beating pattern repeating after $\approx 6-7$ added
charges. A strong beating pattern on the n-side has been observed
before \cite{Park} and has been attributed to the formation of a
small quantum dot in the vicinity of one of the contacts. Since
the metal contact has a larger workfunction than the NT, p-doping
is expected at the contacts. If this charge transfer is large
enough a p-type metal-insulator-semiconductor contact arises. As
the p-type puddle is well buried below the contacts it possibly
cannot be depleted by the gate whose field is screened by the
contacts. Then, the p-type puddle remains even if the bulk NT is
n-type.


The analysis of CB diamonds permits to extract the factor $\alpha$
\cite{Beenakker} which measures the effectiveness of the coupling
capacitance between the tube and the gate, i.e.,
$\alpha=C_g/C_\Sigma=U_c/\Delta V_g$. Here, $C_g$ is the gate
capacitance, $C_\Sigma$ the total capacitance (gate plus
contacts), $U_c = e^2 / C_\Sigma$ the charging energy, and $\Delta
V_g$ the single-electron period in gate voltage. We estimate the
charging energy from the averaged value of the addition energy of
a set of Coulomb diamonds. This results in \mbox{$U_c
\approx$\,2.5 meV} within a gate-voltage period of \mbox{$\Delta
V_g = 12$\,meV}, from which we deduce \mbox{$\alpha\approx 0.2$}.
Note, that this is a very high coupling effectiveness for a
nano\-tube whose gate-electrode is as much as \mbox{$400$\,nm}
away! We take this coupling effectiveness to estimate the size of
the semiconducting gap $E_g$ using the measurement of figure~2 or
figure~3. The SG corresponds to a gate-voltage window of
\mbox{$\Delta V_{g-gap} \approx 4-5$\,V} leading to
\mbox{$E_{gap}=e\alpha\Delta V_{g-gap} = 0.8-1.0$\,eV}. This value
is comparable to the one given above and is in fair agreement with
the reported \mbox{$0.8$\,eV} for a \mbox{$1$\,nm} diameter SWNT
\cite{Dekker2}. From $\alpha$ and $U_c$ we obtain for the
capacitances \mbox{$C_g \approx 12.8$\, aF} and \mbox{$C_\Sigma
\approx 60$\,aF}. $C_g$ is in reasonable agreement with the
estimated geometrical capacitance \mbox{$C_{geometry}= 2\pi
L\epsilon_r\epsilon_{0}/ln[2L/d]$} ($L$ and $d$ are the length and
the radius of the nanotube, respectively), yielding
\mbox{$29$\,aF}. The factor of $2$ difference may originate from
the partial screening by the contacts.

The value of $\alpha$ found here is one of the largest values
reported so far \cite{Radosavljevic}. As a reference to our study
on semiconducting SWNTs, we have investigated metallic tubes of
similar length too. Contrary to semiconducting NTs we observe
regular Coulomb blockade diamonds in metallic SWNTs (see
figure~4c). At the edges of each diamond, parallel and sharp lines
are visible reflecting excited states of the nanotube quantum
dots. The charging energy and the single electron level spacing
are found to be \mbox{$3$\,meV} and \mbox{$1$\,meV}, respectively.
The latter is in good agreement with the contact separation of
\mbox{$L \approx 1$\,$\mu$m}. This strongly suggests, that
metallic SWNTs behave like single quantum dots, unlike
semiconducting SWNTs. In addition, the coupling to the gate shows
the same large value of \mbox{$\alpha \sim 0.2-0.3$} and
corroborate the universal aspect of the high gate effectiveness
independent of the nature of the tubes. We think that the use of
hydrogen during the CVD growth is of crucial importance. It
passivates SiO$_2$ dangling bonds and helps to reduce the number
of charge traps in the substrate. Consequently, it leads to a
better coupling of SWNT to the back-gate and therefore to bipolar
action of the devices.

In conclusion, these experiments demonstrate that CVD-grown SWNTs
can display a very high coupling efficiencies to a back gate
without using any additional post-treatment. Due to the large
coupling efficiency semiconducting SWNTs can be continuously gated
from p to n-side and therefore act as bipolar FETs. At
low-temperature, semiconducting SWNTs are strongly affected by
disorder, which (partially) splits the tube in a couple of quantum
dots. As the maximum number of dots is smaller than $3$ for our
SWNTs with \mbox{$1$\,$\mu$m} contact separation, it appears that
single quantum dots should be feasible for smaller contact
separation.

 \begin{acknowledgments}
We acknowledge M.R. Buitelaar for fruitful discussions and T.
Nussbaumer for technical assistance. This work is supported by
COST (BBW), the NCCR on Nanoscience and the Swiss NFS.
 \end{acknowledgments}


\begin{figure}
\begin{center}
\includegraphics[width=120mm]{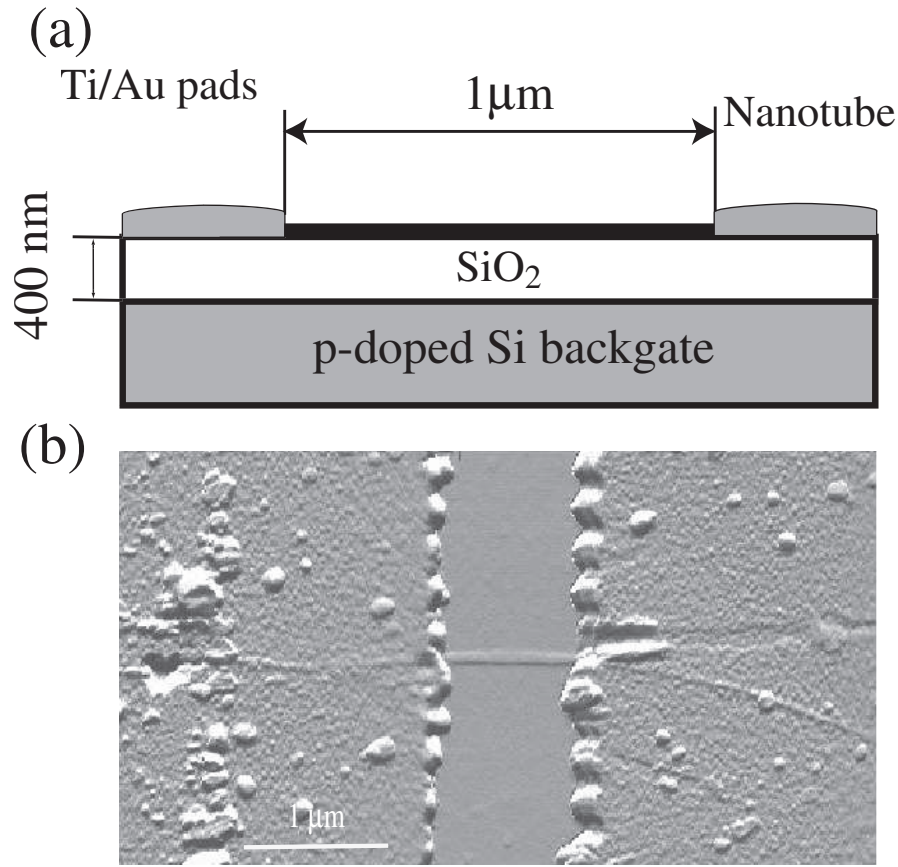}
\end{center}
\caption{\label{Figure 1.} (a) Scheme of a SWNT device contacted
by two Ti/Au electrodes.  The Si substrate is used as back-gate.
(b) Atomic force microscopy (AFM) image of a SWNT bridging between
the two electrodes.}

\end{figure}

\begin{figure}
\begin{center}
\includegraphics[width=120mm]{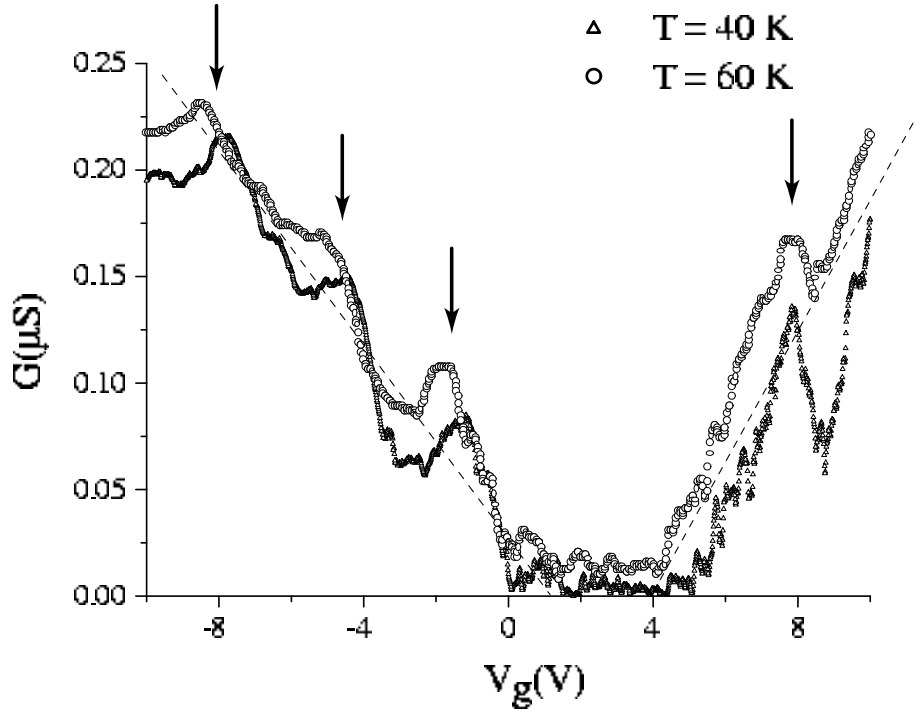}
\end{center}
\caption{\label{Fig. 2.} Two-terminal conductance $G$ as a
function of gate-voltage $V_g$ for a semiconducting SWNT at
moderate temperatures of \mbox{$T=40$} and \mbox{$60$\,K},
respectively. The peaks in $G$ (arrows) are attributed to van Hove
singularities in the DOS.}
\end{figure}

\begin{figure}
\begin{center}
\includegraphics[width=120mm]{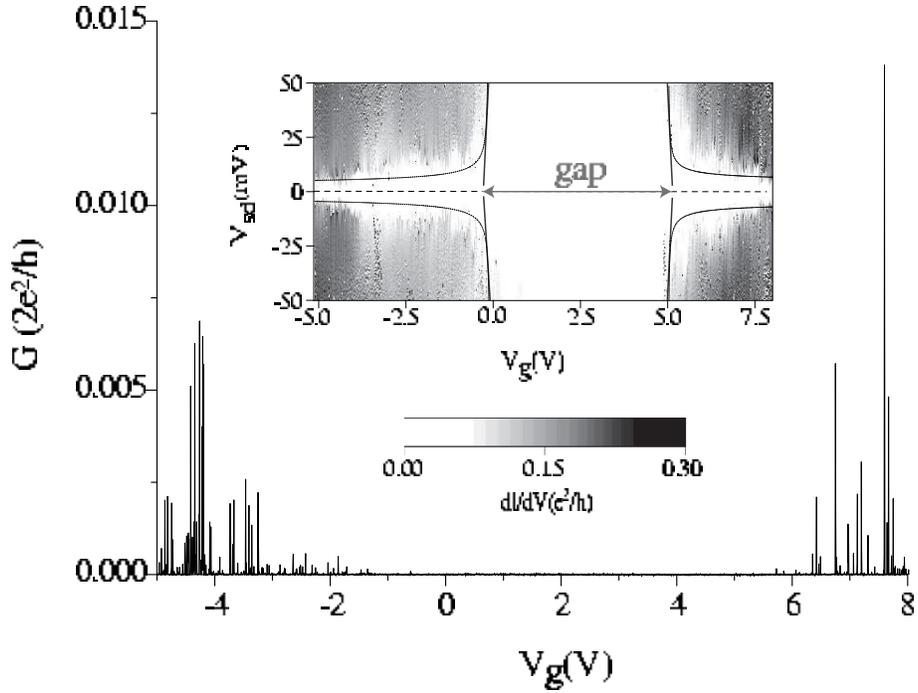}
\end{center}
\caption{\label{Fig. 3.} Linear conductance $G$ as a function of
gate voltage $V_g$ (main plot) and greyscale representation of the
differential conductance $dI/dV$ as a function of $V_g$ and
applied source-drain voltage $V_{sd}$ (inset) at  \mbox{$2$\,K}
for a SWNT device. White regions correspond to zero and dark
regions to high conductances (maximum \mbox{$0.3$\, $e^2/h$)}. The
semiconducting gap (SG) is clearly visible as a large
non-conducting region in the inset. Coulomb oscillations peaks are
observed on the p (left) and n (right) side of the semiconducting
gap.}
\end{figure}

\begin{figure}
\begin{center}
\includegraphics[width=120mm]{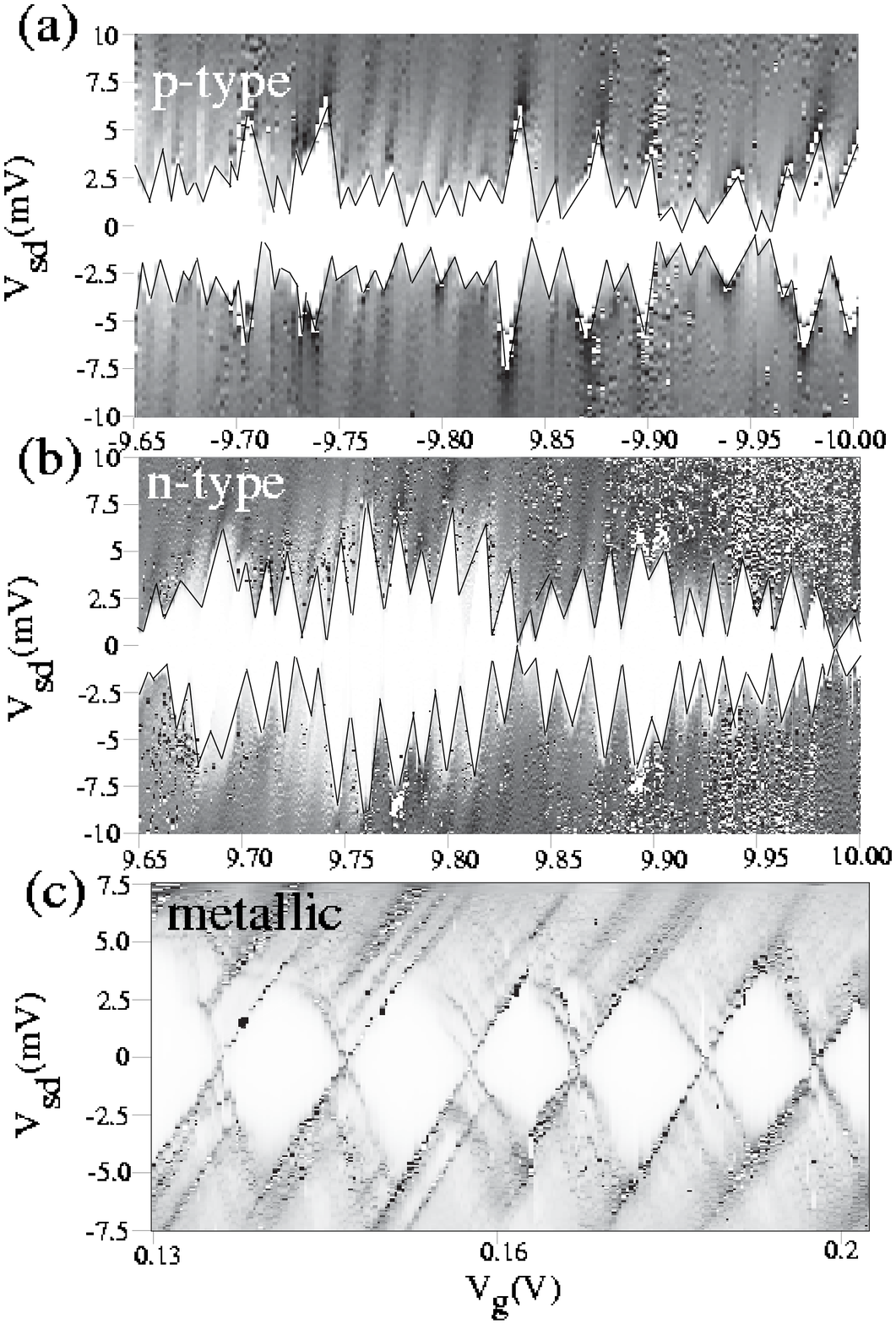}
\end{center}
\caption{\label{Fig. 4.}Differential conductance ($dI/dV$) plots
as a function of $V_g$ and $V_{sd}$. White corresponds to
$dI/dV=0$, and black to the maximum conductance of
\mbox{$0.3$\,$e^2/h$}. (a) and (b) have been measured on one
semiconducting SWNT in the p region (a) and n region (b) at
\mbox{$2$\,K}. As a reference a similar plot of another metallic
SWNT measured at \mbox{$0.3$\,K} is shown in (c). While excited
states can clearly been seen in (c), they appear to be absent in
(a) and (b). Furthermore, while (c) displays a regular Coulomb
blockade (CB), it is irregular in (a) and (b). In all cases strong
coupling to the gate is inferred.}
\end{figure}

\begin{thebibliography}{99}

\bibitem{Dresselhaus}
See e.g., Dresselhaus M S, Dresselhaus G, Eklund P C 1996: {\it
Science of Fullerenes and Carbon Nanotubes (New York, Academic
Press, 1996)}

\bibitem{Dekker1}
Dekker C 1999 {\it Physics Today} {\bf 52} (5) 22

\bibitem{Dai2}
Kong J {\em et al} 2000 {\it Phys. Rev. Lett.} {\bf 87} 106801

\bibitem{mcuene1}
Bockrath M {\em et al} 1999, {\it Nature} {\bf 397} 598

\bibitem{Liang}
Liang W {\em et al} 2001 {\it Nature} {\bf 411} 665

\bibitem{Tans1}
Tans S J {\em et al} 1998 {\it Nature} {\bf 393} 49

\bibitem{Martel1}
Martel R, Schmidt T, Hertel T, and Avouris P 1998 {\it Appl. Phys.
Lett.} {\bf 73} 2447

\bibitem{Dai1}
Zhou C {\em et al} 2000  {\it Science} {\bf 290} 1552

\bibitem{Park}
Park J and McEuen P 2001 {\it Appl. Phys. Lett}  {\bf 79} 1363

\bibitem{Bockrath}
Bockrath M {\em et al} 2000 {\it Phys. Rev. B} {\bf 61} R10606

\bibitem{Radosavljevic}
Radosavljevi\'c M, Feitag M, Thadani K V and Johnson A T 2002 {\it
Nano Lett.} {\bf 2} No. 7 p 761

\bibitem{Javey}
Javey A, Shim M and Dai H 2002 {\it Appl. Phys. Lett} {\bf 80}
1064

\bibitem{Bachtold}
A Bachtold, P Hadley, T Nakanishi and C Dekker 2001 {\it Science}
{\bf294} 1317

\bibitem{Hafner1}
Hafner J H, Bronikowski M J, Azamian B R, Nikolaey P, Rinzler A G,
Rinzler D T, Smith K A and Smalley R E 1998 {\it Chem. Phys.
Lett.} {\bf 296} 195

\bibitem{us1}
From gate period for CB peaks we add approximately $100$ electron
per one Volt at the gate.

\bibitem{Beenakker}
Beenakker C W J 1991 {\it Phys. Rev.} {\bf 44} 1646

\bibitem{Dekker2}
Wildoer J  {\em et al} 1998 {\it Nature} {\bf 391} 59



\end{thebibliography}
\end{document}